\documentclass[%
 reprint,
 amsmath,amssymb,
 aps,
]{revtex4-1}
\usepackage{blindtext}
\usepackage{comment}
\usepackage{subfigure}
\usepackage{graphicx}
\usepackage{dcolumn}
\usepackage{bm}
\usepackage{amssymb}
\usepackage{float}
\usepackage{graphicx}
\usepackage{mathtools}
\usepackage{type1cm}
\usepackage{eso-pic}
\usepackage{color}
\usepackage[scientific-notation=true]{siunitx}
\begin{document}
\title{
Two-scale evolution during shear reversal in dense suspensions}
\author{Christopher Ness}
\author{Jin Sun}
\affiliation{School of Engineering, University of Edinburgh, Edinburgh, EH9 3JL, United Kingdom}
\date{\today}
\begin{abstract}

We use shear reversal simulations to explore the rheology of dense, non-Brownian suspensions, resolving lubrication forces between neighbouring particles and modelling particle surface contacts. The transient stress response to an abrupt reversal of the direction of shear shows rate-independent, nonmonotonic behaviour, capturing the salient features of the corresponding classical experiments. 
Based on analyses of the hydrodynamic and particle contact stresses and related contact networks, we demonstrate distinct responses at small and large strains, associated with contact breakage and structural re-orientation, respectively, emphasising the importance of particle contacts.
Consequently, the hydrodynamic and contact stresses evolve over disparate strain scales and with opposite trends, resulting in nonmonotonic behaviour when combined.
We further elucidate the roles of particle roughness and repulsion in determining the microstructure and hence the stress response at each scale.

\end{abstract}
\maketitle

\section{Introduction}

The flow behaviour of dense suspensions is strongly affected by details of the microstructure and interparticle forces~\cite{Mewis2011}. Recent theoretical~\cite{Wyart2014}, experimental~\cite{Fernandez2013,Guy2015} and computational~\cite{Seto2013,Mari2014a} work suggests that particle surface contacts make a major contribution to suspension rheology, though their precise role, and importance relative to hydrodynamic interactions, is still debated. To this end, shear reversal experiments, in which the flow direction is suddenly reversed, prove to be an elegant means of estimating the contact stress (CS) contribution, while probing microstructural anisotropy. For example, it was shown that following a flow cessation period, the shear~\cite{Acrivos1980} (or normal~\cite{Kolli2002a}) stress in a suspension of $\sim 40 \ \mathrm{\mu m}$ polystyrene spheres reaches an ``immediate'' peak upon reversal, then evolves nonmonotonically over a strain of $\approx3$ to its steady state. The large-strain evolution was attributed to microstructural realignment~\cite{Acrivos1980}; the initial stress peak was hypothesized to represent the hydrodynamic stress (HS), leading to a suggestion of the larger role played by particle contacts in denser suspensions~\cite{Kolli2002a}. It remains difficult to isolate the evolution of the CS and HS contributions and link the microstructural effect to the puzzling nonmonotonic behaviour.
In the present paper, we reveal responses at two disparate strain scales, an elegent manifestation of the microfragile vs. macrofragile distinction proposed by Cates~et~al.~\cite{Cates1998}. The small strain stress peak is shown to be a hydrodynamic response to surface-contact breakage, but is distinct from the steady state HS. The large strain scale is determined by microstructural reorientation, as predicted \cite{Acrivos1980}. The nonmonotonic behaviour is a combined effect of the CS and HS evolutions. We show that different surface characteristics control the stress response at different strain scales, meaning our two-scale explanation, and hence the micro- versus macro- fragility paradigm, can be applied usefully to a wide range of suspended systems.

\section{Simulation model}

We solve the equations of motion numerically \cite{Plimpton1995} for neutrally buoyant suspended non-Brownian particles, subject to forces and torques arising due to hydrodynamics and particle surface contact \cite{Cundall1979}. For dense suspensions, in which the average surface separation between neighbouring particles becomes very small, the full hydrodynamic resistance matrix \cite{Brady1988} can be suitably approximated by resolving pairwise, frame-invariant lubrication forces \cite{Ball1997}, which diverge at contact and significantly exceed the long-range force components. Such a simplification has been proven to be effective in capturing the behaviour of dense suspensions~\cite{Trulsson2012,Mari2014a,Kumar2010}.
For an interaction between particles $i$ and $j$, (with particle and fluid density $\rho$) the force and torque on particle $i$ due to hydrodynamic lubrication can be expressed as
\begin{subequations}
\begin{align}
&
\begin{multlined}
\mathbf{F}^{l}_{ij} = -a_{sq} 6 \pi \eta_f (\textbf{v}_i - \textbf{v}_j) \cdot \textbf{n}_{ij} \textbf{n}_{ij}\\
- a_{sh} 6 \pi \eta_f (\textbf{v}_i - \textbf{v}_j) \cdot (\mathbf{I}-\mathbf{n}_{ij}\mathbf{n}_{ij}) \text{,}
\end{multlined}\\
&
\begin{multlined}
\mathbf{\Gamma}^{l}_{ij} = -a_{pu} \pi \eta_f d_i^3(\boldsymbol{\omega}_i - \boldsymbol{\omega}_j) \cdot (\mathbf{I}-\mathbf{n}_{ij}\mathbf{n}_{ij})\\
- \frac{d_i}{2} \left(\mathbf{n}_{ij} \times \mathbf{F}^{l}_{ij}\right) \text{,}
\end{multlined}
\end{align}
\label{eq:lube_forces}
\end{subequations}
for particle diameter $d_i$, fluid viscosity $\eta_f$, particle translational and rotational velocity vectors $\mathbf{v}_i$ and $\boldsymbol{\omega}_i$ respectively, centre-to-centre unit vector $\mathbf{n}_{ij}$ pointing from particle $j$ to $i$ and identity tensor $\mathbf{I}$, with the squeeze $a_{sq}$, shear $a_{sh}$ and pump $a_{pu}$ resistance terms as derived by \cite{Kim1991}, for $\beta = d_j/d_i$, as:
\begin{subequations}
\begin{align}
&
\begin{multlined}
a_{sq} = \frac{\beta^2}{(1+\beta)^2} \frac{d_i}{2h}	+	\frac{1 + 7\beta + \beta^2}{5(1 + \beta)^3} \frac{d_i}{2} \ln \left(\frac{d_i}{2h} \right)\\
+ \frac{1 + 18\beta - 29\beta^2 + 18\beta^3 + \beta^4}{21(1+\beta)^4} \frac{d_i^2}{4h} \ln \left(\frac{d_i}{2h}\right) \text{,}
\end{multlined}\\
&
\begin{multlined}
a_{sh} = 4 \beta \frac{2 + \beta + 2\beta^2}{15 (1 + \beta)^3} \frac{d_i}{2} \ln \left( \frac{d_i}{2h}\right)\\
+ 4\frac{16 -45\beta + 58\beta^2 - 45\beta^3 + 16\beta^4}{375(1 + \beta)^4} \frac{d_i^2}{4h} \ln \left( \frac{d_i}{2h} \right) \text{,}
\end{multlined}\\
&
\begin{multlined}
a_{pu} = \beta \frac{4 + \beta}{10(1 + \beta)^2} \ln \left( \frac{d_i}{2h} \right)\\
\frac{32 - 33\beta + 83\beta^2 + 43\beta^3}{250\beta^3} \frac{d_i}{2h} \ln \left( \frac{d_i}{2h} \right) \text{.}
\end{multlined}
\end{align}
\label{eq:resistances}
\end{subequations}
 The separation between particles $i$ and $j$ is calculated according to $h = |\mathbf{r}_{ij}| - \frac{d_i + d_j}{2}$ for centre-to-centre vector $\mathbf{r}$. We calculate the lubrication force when the interparticle gap $h$ is smaller than $h_\text{max} = 0.05 d$ (where $d$ is the harmonic average particle diameter).
An increasing body of evidence \cite{Fernandez2013,Guy2015} shows that direct particle-particle surface contacts can play a major role in suspension rheology; indeed, simulations that strictly resolve lubrication forces (treating particles as ideally hard and ideally smooth)~\cite{Bossis1989a} have proven to be inadequate for capturing dense suspension rheology for cases where particle-particle contacts are presumed to be important. Therefore we truncate the lubrication divergence and regularize the contact singularity at a typical asperity length scale $h_\text{min}$ ($=0.001d$ unless specified otherwise), i.e., setting $ h = h_\text{min}$ in the force calculation, when $h < h_\text{min}$. We use a value of $\eta_f = 0.1$ [viscosity unit: $\rho d^2/t$].



Mechanical contact occurs at $h\leq0$, giving normal repulsive and tangential forces described by a linear spring model and related through a Coulomb friction coefficient $\mu_p$ ($=0.2$ unless specified otherwise) \cite{Cundall1979}. A linear (as opposed to Hertzian) spring is chosen for convenience, though we expect Hertzian results to lead to identical conclusions reagarding the respective roles of contacts and lubrication.
The normal ($\mathbf{F}^{c,n}$) and tangential ($\mathbf{F}^{c,t}$) contact force and torque $\mathbf{\Gamma}^c$ are given by
\begin{subequations}
\begin{equation}
\mathbf{F}^{c, n}_{ij} = k_\text{n} \delta \mathbf{n}_\text{ij} \text{,}
\end{equation}
\begin{equation}
\mathbf{F}^{c,t}_{ij} = -k_\text{t} \mathbf{u}_\text{ij} \text{,}
\end{equation}
\begin{equation}
\mathbf{\Gamma}^{c}_{ij} = -\frac{d_i}{2} (\mathbf{n}_{ij} \times \mathbf{F}^{c,t}_{ij}) \text{,}
\end{equation}
\end{subequations}
for a collision between particles $i$ and $j$ with normal and tangential spring stiffnesses $k_n$ and $k_t$ respectively [$k_n = 20000$, unit: $\rho d^3/t^2$ and $k_t = (2/7)k_n$], particle overlap $\delta$ and tangential displacement $\mathbf{u}_\text{ij}$.

The bulk stress tensor is calculated from the particle force and velocity data. It is decomposed into contributions due to the hydrodynamic interaction and the particle-particle interaction, given by Eqs.~\ref{eq:stressF}~and~\ref{eq:stressC}, respectively,  
\begin{subequations}
\begin{align}
\bm{\sigma}^l_{ij} &= \frac{1}{V} \sum_i \sum_{i \neq j} \mathbf{r}_{ij} \mathbf{F}^l_{ij} \text{,} 
\label{eq:stressF} \\
\bm{\sigma}^c_{ij} &= \frac{1}{V} \sum_i \sum_{i \neq j} \mathbf{r}_{ij} (\mathbf{F}^{c,n}_{ij} + \mathbf{F}^{c,t}_{ij}) \text{.}
\label{eq:stressC}
\end{align}
\end{subequations}
In the following discussion, we consider the shear components of the above stress tensors corresponding to the direction of the applied deformation, $\sigma^l$ and $\sigma^c$, as well as the mean of the diagonal components, namely the ``pressures'' $P^l$ and $P^c$.
The hydrodynamic stress $\sigma^l$ is further decomposed in two ways. In the first, we isolate the contributions from normal forces (the squeeze $a_{sq}$ terms) and tangential forces (the shear $a_{sh}$ and pump $a_{pu}$ terms) \cite{Ball1997} as $\sigma^l_\text{normal}$ and $\sigma^l_\text{tangential}$ respectively. In the second, we isolate contributions from opening and closing particle pairs (pairs for which $dh/dt>0$ and $dh/dt<0$ respectively), presented as $\sigma^l_\text{opening}$ and $\sigma^l_\text{closing}$. It is noted that $\sigma^l_\text{opening}+\sigma^l_\text{closing} = \sigma^l$ and $\sigma^l_\text{normal}+\sigma^l_\text{tangential} = \sigma^l$.
Assemblies of 5000 spheres are sufficiently large to achieve system size independence, and bidispersity with diameter ratio $1:1.4$ prevents crystallisation \cite{Ikeda2012}.
Simulation results are ensemble averaged over 20 realizations with different initial particle configurations.
We note that although the overlap is exceedingly small, typically of order $10^{-7}d$ in the Stokesian regime~\cite{Ness2015}, it can lead to qualitatively different rheology from that produced using the ``ideal'' hard-sphere model, as demonstrated later.
The present technique produces results at the dense limit (solid volume fraction $\phi \gtrsim 0.45$) closely approximating those that would be obtained by fully resolving the hydrodynamics (e.g. \cite{Brady1985}), but assuming particles co-move with fluid at the mean flow level~\cite{Ball1997}, valid for shear flows. We verify this by incorporating an additional drag force, similar to~\cite{Trulsson2012}, which leads to a negligible increase in the calculated suspension viscosity. 
The particle assemblies are subjected to rate ($\dot{\gamma}$) controlled simple shear flow in a 3-dimensional periodic domain at constant $\phi$ ($= 0.54$) and Stokes numbers $\text{St}$~($=\rho \dot{\gamma}d^2/\eta_f$)~$<10^{-2}$, inhibiting particle inertia. The suspension is first sheared from $\dot{\gamma}t = -8 \to -2$, reaching steady flow. 
No shear is applied for $\dot{\gamma}t = -2 \to 0$.
From $\dot{\gamma}t = 0$, the suspension is sheared in the opposite direction until a new steady state is obtained.

\section{Stress and microstructure evolution}

\begin{figure}
\centering
      \subfigure[c][]{
  \includegraphics[width=76mm]{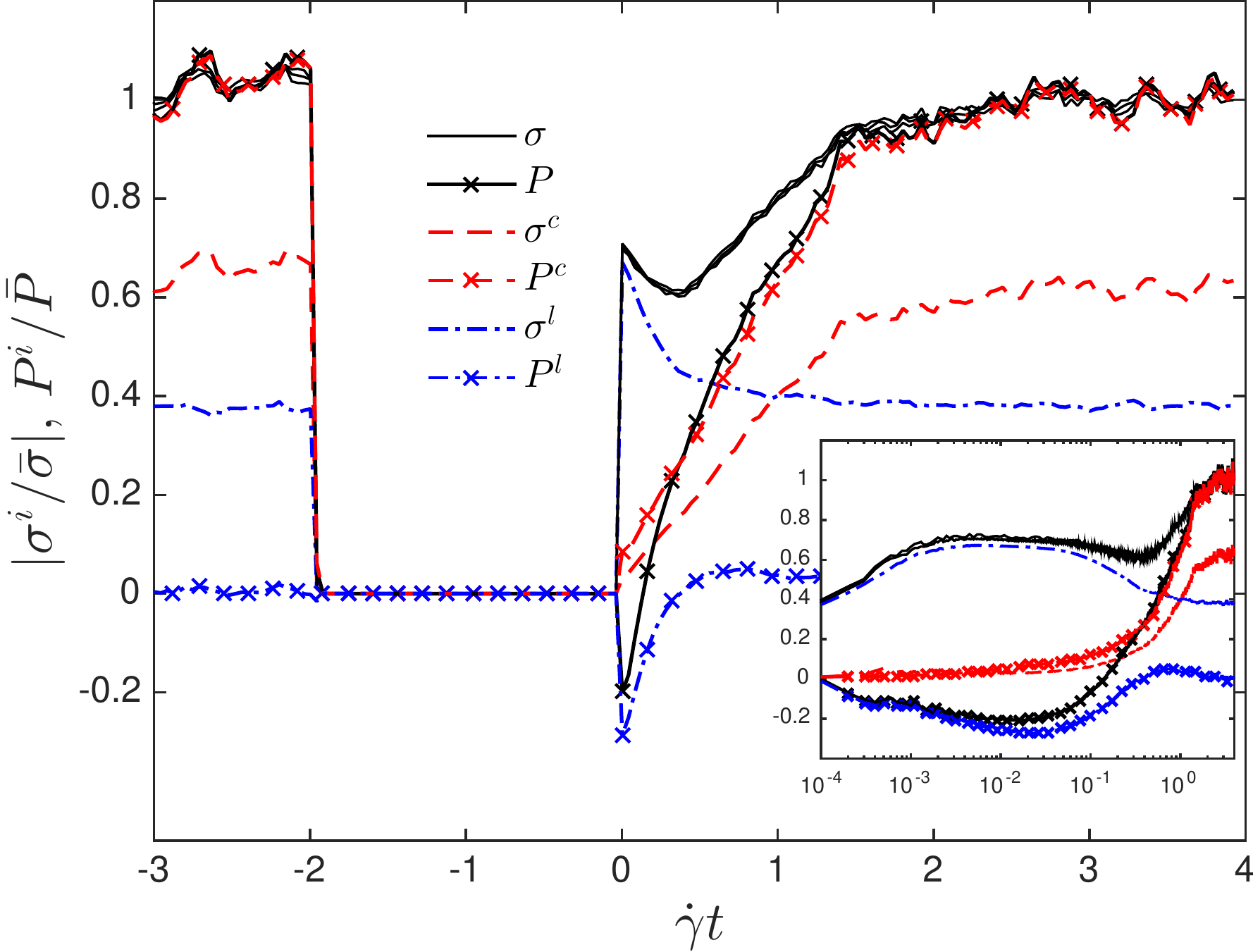}}
          \subfigure[]{
  \includegraphics[width=75mm]{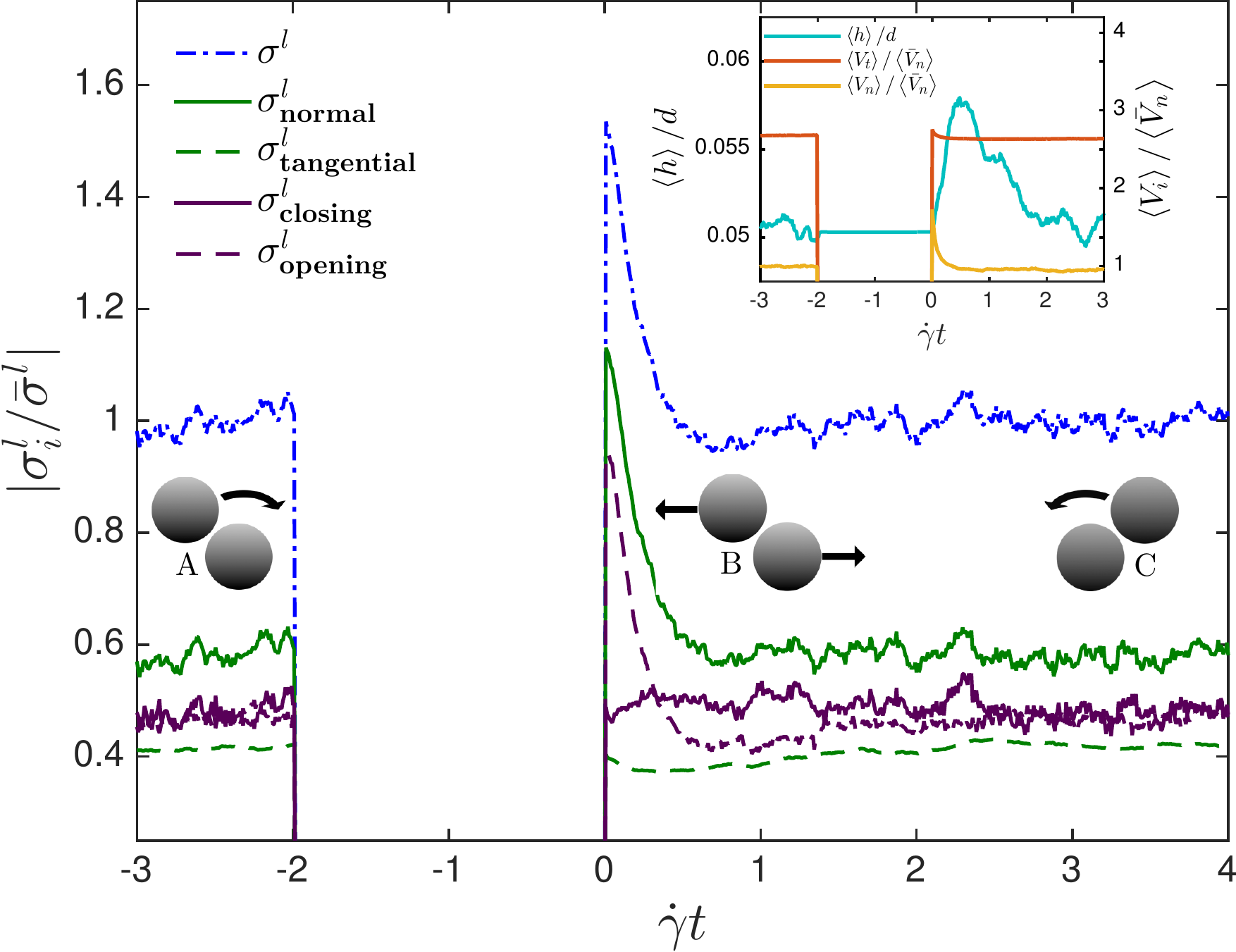}}
            \subfigure[]{
\hspace{6mm}  \includegraphics[width=80mm]{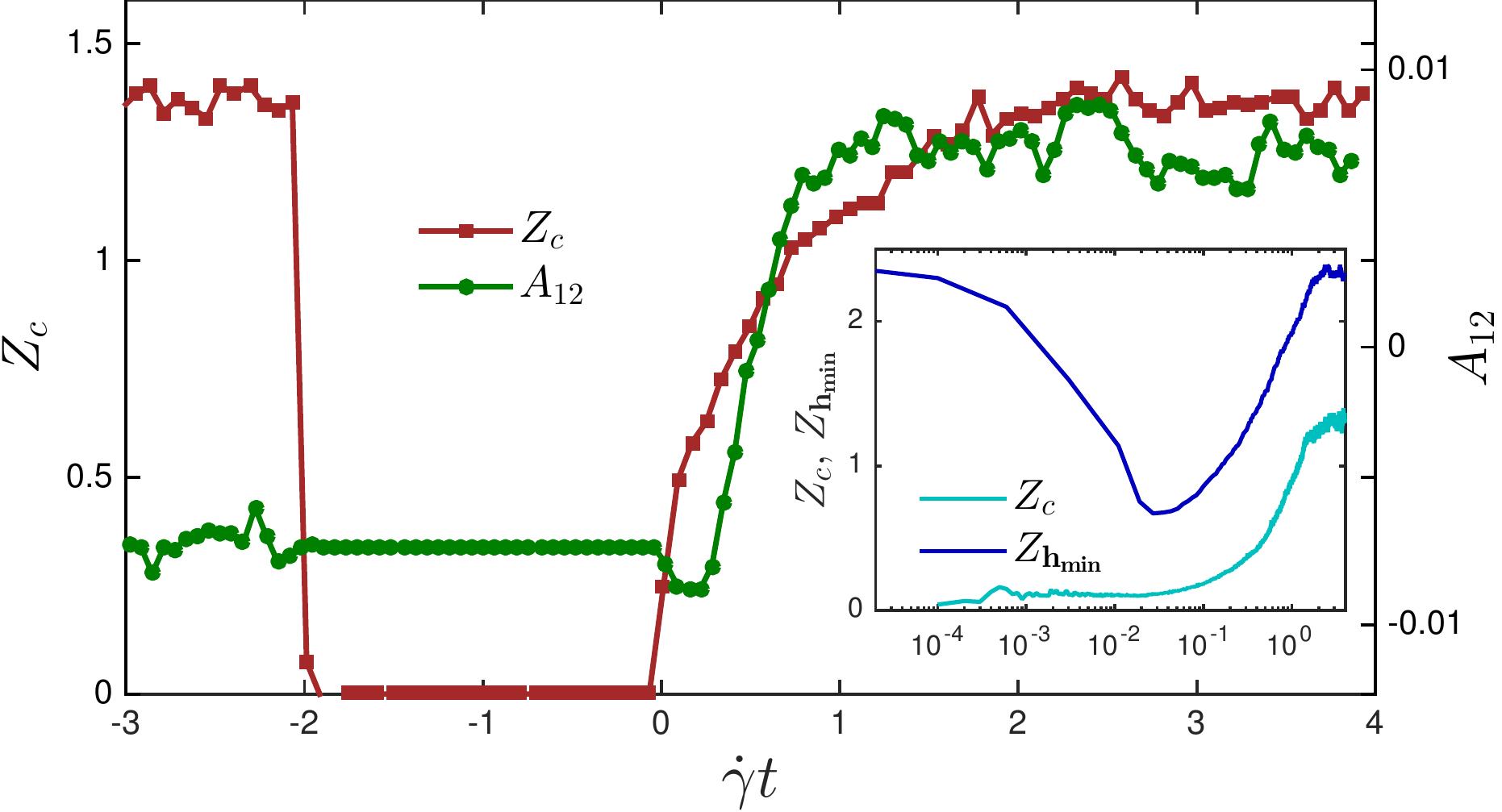}}
          \caption{
          (a) Evolution of stresses following reversal at $\dot{\gamma}t = 0$, showing total ($\sigma$, $P$), contact ($\sigma^c$, $P^c$) and fluid ($\sigma^l$, $P^l$) contributions, each scaled by the steady state total stress ($\bar{\sigma}$, $\bar{P}$). Multiple lines for $\sigma$ illustrate $\dot{\gamma}$ independence. Inset: Same data with logarithmic $x$-axis;
          (b) $\sigma^l$ components arising from (i) normal and tangential forces; (ii) opening and closing interparticle gaps. Inset: Evolution of the mean fluid film thickness $\left<h\right>$ and scaled mean neighbouring-particle normal $\left< V_n \right>/\left< \bar{V}_n \right>$ and tangential $\left< V_t \right>/\left< \bar{V}_n \right>$ velocity magnitudes. Embedded: particle-pair configurations corresponding to different times.
          (c) Evolution of coordination number $Z_c$ and shear fabric component $A_{12}$. Inset: Coordination $Z_c$ and surface coordination $Z_{\mathrm{h_{min}}}$ evolution, with logarithmic $x$-axis.
}
\label{fig:base_case}
\end{figure}

 The total stress ($\sigma = \sigma^c + \sigma^l$) evolution, Fig~\ref{fig:base_case}(a), is strikingly reminiscent of classical experiments~\cite{Acrivos1980,Narumi2002a,Kolli2002a}. Rate-independence is demonstrated by collapsing stress components with the respective steady state total stress~$\bar{\sigma}$, for multiple $\dot{\gamma}$.
Fig.~\ref{fig:base_case}(c) shows the microstructural evolution, characterised by a mechanical coordination number $Z_c$, the mean number of per particle contacts which support a contact stress greater than $10^{-6}$ of the mean steady-state stress $\bar{P}$, a surface coordination number $Z_{\mathrm{h_{min}}}$ counting all pairs with $h < h_\text{min}$ and a fabric tensor~\cite{Sun2011,Ness2015}, $\mathbf{A}=2/(Z_{\mathrm{h_{max}}}N) \sum_{h<h_\mathrm{max}} \mathbf{n}_{ij} \mathbf{n}_{ij}-\frac{1}{3}\mathbf{I}$. Under shear flow, particles preferentially align along the compressive axis at $45^{\circ}$  with the shear component of $\mathbf{A}$, $|A_{12}| = 0.5$ representing perfect alignment of all contacts and $A_{12} = 0$ representing perfect isotropy. The evolution of additional microstructure variables is shown in Appendix~\ref{appendix:2}, Fig~\ref{fig:base_case_addition}. 

\subsection{Steady flow and cessation}

In the steady state ($\dot{\gamma}t<-2$, Fig~\ref{fig:base_case}(a)), the contact stress contribution is surprisingly large given the small contact overlaps, representing 60\% of the total shear stress. The relative contribution is $\phi$-dependent, e.g., at $\phi=0.47$, we found $\sigma^c \approx 0.3 \sigma$~\cite{Ness2015}. We find $Z_c \approx1.5$ and the shear component of $\mathbf{A}$, $A_{12} \approx -0.01$, indicating persistent mechanical contacts and an anisotropic network of lubrication films. In this condition, relative particle motions are as illustrated in particle-pair diagram A, Fig~\ref{fig:base_case}(b), a configuration that results in the mean relative normal velocity of neighbouring particles $\left< V_n \right>$ being smaller than the mean relative tangential velocity $\left< V_t \right>$, highlighted in Fig~\ref{fig:base_case}(b) Inset, which gives these quantities scaled by the steady state value of $\left< V_n \right>$, $\left< \bar{V}_n \right>$. This leads to comparable normal and tangential lubrication forces (and corresponding stress contributions $\sigma^l_\text{normal}$ and $\sigma^l_\text{tangential}$ as decomposed in Fig.~\ref{fig:base_case}(b)), in spite of the order of magnitude difference expected from their respective $1/h$ versus $\ln(1/h)$ dependence. The average number of particle pairs moving together or apart is equal at steady state, as required to satisfy the constant volume constraint, resulting in constant mean lubrication film thickness $\left<h \right>$, Fig.~\ref{fig:base_case}(b) Inset, equal stress contributions $\sigma^l_\text{opening}$ and $\sigma^l_\text{closing}$, Fig.~\ref{fig:base_case}(b), and $\bar{P}$ dominated by $P^c$, Fig~\ref{fig:base_case}(a).
 
Upon flow cessation ($\dot{\gamma}t = -2$), the contact stress relaxes together with the hydrodynamics stress, suggesting that caution should be exercised when interpreting the ``instantaneous'' stress loss as entirely hydrodynamic in such experiments \cite{Mewis2011,OBrien2000}. 
Correspondingly, $Z_c$ drops to zero in the relaxation period, though a small portion of weak contacts relax more slowly due to confinement and fluid overdamping. The shear-induced anisotropic microstructure pertaining to hydrodynamics, however, remains intact throughout the relaxation period, evidenced by constant $A_{12}$ and $Z_{\mathrm{h_{min}}}$ (Fig~\ref{fig:base_case}(c) and Inset, $Z_\mathrm{h_{min}}(\dot{\gamma}t\to0) = Z_\mathrm{h_{min}} (\dot{\gamma}t = 4)$), implying the steady state HS can be recovered instantaneously (with opposite sign) upon shear reversal.

\subsection{Shear reversal: micro- and macro-strain responses}

Indeed, $\sigma^l$ does resume it's steady state magnitude upon reversal for strains $\leq10^{-4}$, Fig.~\ref{fig:base_case}(a) Inset. It then surges to a significant peak, around 50\% greater than the steady value, at strain $10^{-3}$, sustaining until about $10^{-2}$ where it starts to subside. Resumption of the steady value followed by a demonstrable peak is also observed for $P^l$ over the same strain scale. We attribute this small-strain surge, the manifestation of a microfragile response \cite{Cates1998}, to the pulling apart of particle surfaces at the $h_\mathrm{min}$ ($=10^{-3}d$) scale due to the new (reversed) load being incompatible with the present microstructural alignment. This is clearly demonstrated by the coincident decrease of $Z_{\mathrm{h_{min}}}$, which reaches a minimum near $10^{-2}$.  The mechanism is further evidenced by the significantly greater $\left< V_n \right>$ than under steady flow, the dominance of $\sigma^l_\text{normal}$ and $\sigma^l_\text{opening}$, and the tensile nature of $P^l$. This is further illustrated by the evolution of the $h$ distribution, given in Appendix~\ref{appendix:1}. Relative particle motions during this time are illustrated in particle-pair diagram B, Fig.~\ref{fig:base_case}(b). Such microfragile events in, e.g., a dry granular system, would be subtle to detect or difficult to distinguish from the macroscopic process. These events in dense suspensions, \emph{nonhydrodynamic} in nature, however, lead to the spectacular \emph{hydodynamic} responses, which have been measured robustly in experiments~\cite{Acrivos1980,Kolli2002a}. We note that a microfragile hydrodynamic response is absent in Stokesian Dynamics simulations of shear reversal that strictly inhibit fluid films smaller than $0.01d$ \cite{Bricker2007}, strengthening the argument for direct surface contacts \emph{in addition} to hydrodynamics, as a crucial contributor to the rheology observed by \cite{Acrivos1980,Kolli2002a}.

The subsequent building up of $Z_{\mathrm{h_{min}}}$ after a strain of $0.01$ is coupled to re-orientation of the microstructural anisotropy $A_{12}$, corresponding to macrofragile evolution at a larger strain scale of order unity. 
The initial subsidence of $\sigma^l$ from its peak until $ \dot{\gamma}t\approx0.5$ (while $A_{12}< 0$), corresponds to a net opening of lubrication films (see $\left< h \right>$ and $\sigma_\text{opening}$, Fig~\ref{fig:base_case}(b)), consistent with the leading $1/h$ dependence of the lubrication forces, combined with a reduction in $\left< V_n \right>$.   
At larger strains, a new contact network establishes in the now-compatible, oppositely aligned, compressive direction (evidenced by $A_{12} > 0$) with net repulsive lubrication forces during $0.5<\dot{\gamma}t<2$, restoring $\left< h \right>$ to its steady value thereby producing positive $P^l$ and a marginally dominant $\sigma^l_\text{closing}$. The consequent mean relative particle motion is highlighted in paritcle-pair diagram C, Fig~\ref{fig:base_case}(b). Although $\sigma^l$ evolves continuously during this large-scale period, the responsible mechanism therefore switches as the anisotropy changes sign. The stress presented by~\cite{Bricker2007} has a comparable macrofragile evolution, but exhibits nonmonotonic behaviour due to the absence of a microfragile response.

The contact stresses ($\sigma^c$, $P^c$) follow a similar macrofragile evolution, their associated microfragile contact breakage having occurred at flow cessation as discussed. The stress evolution is closely correlated with the building up of the mechanical coordination number $Z_c$, which occurs on a similar strain scale as the fabric reorientation described above, as illustrated in Fig~\ref{fig:base_case}(c). The separation of scales in the evolution of the contact stress and the hydrodynamic peak ensures dominance of the hydrodynamic stress at small strains after reversal, an assertion made in~\cite{Acrivos1980,Kolli2002a}, though overlooking the microfragile hydrodynamic response. Combining the increasing $\sigma^c$ with the decreasing $\sigma^l$ at $\dot{\gamma}t > 0.01$ gives rise to the nonmonotonic total stress, meaning microfragility in the hydrodynamic response is crucial in capturing the experimental behaviour.

The above analysis sheds light on the two-scale nature of the stress evolution, linked to configurational change at small strains and anisotropy re-orientation at large strains. The importance of particle contacts in achieving the nonmonotonic stress response naturally leads to the question of the sensitivity of the evolution at each scale to particle interactions and surface properties.

\section{Role of particle properties}

\begin{figure}
\subfigure[c][]{
  \includegraphics[width=80mm]{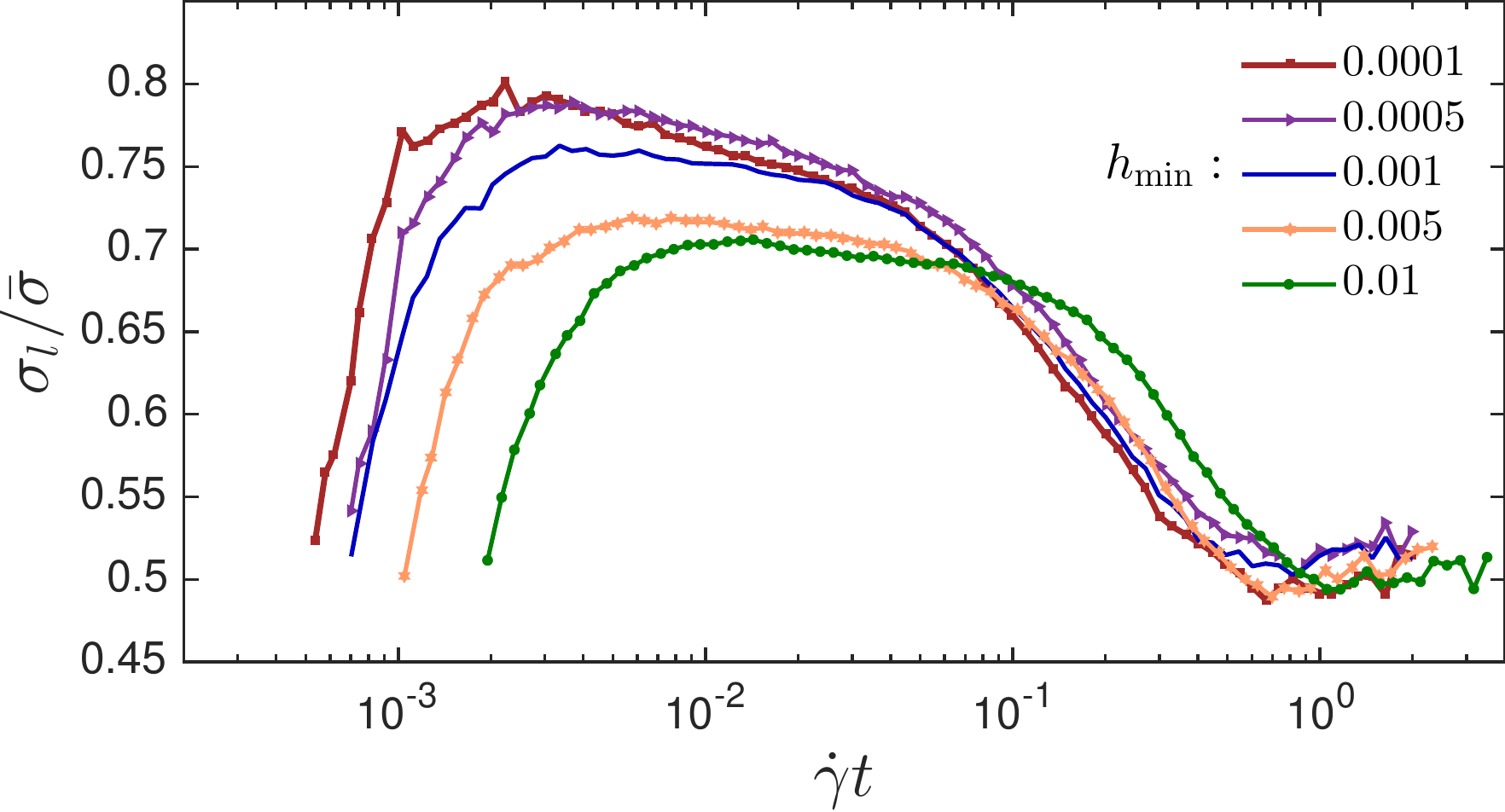}}
          \subfigure[]{
  \includegraphics[width=80mm]{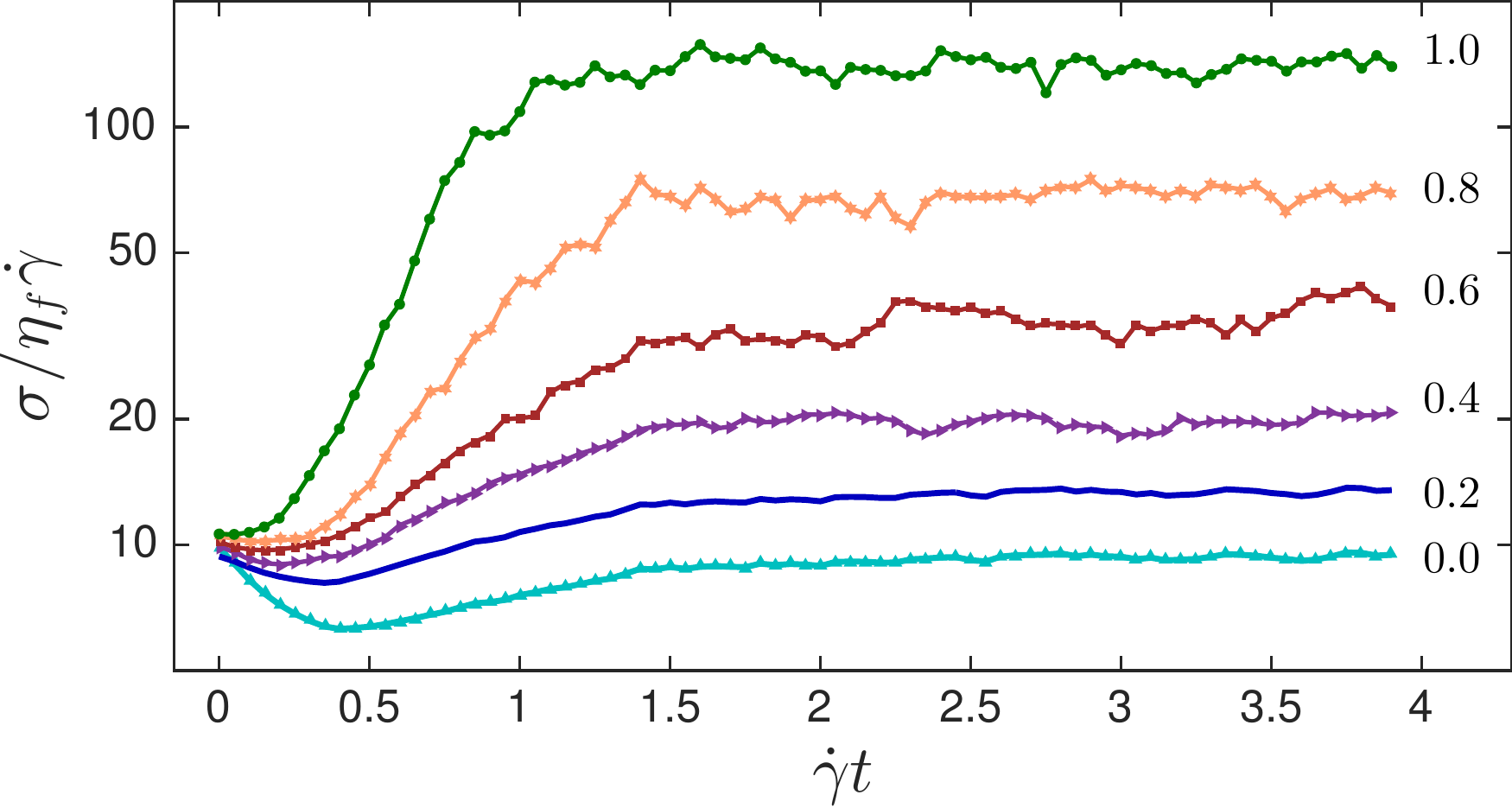}}
            \subfigure[]{
\includegraphics[width=80mm]{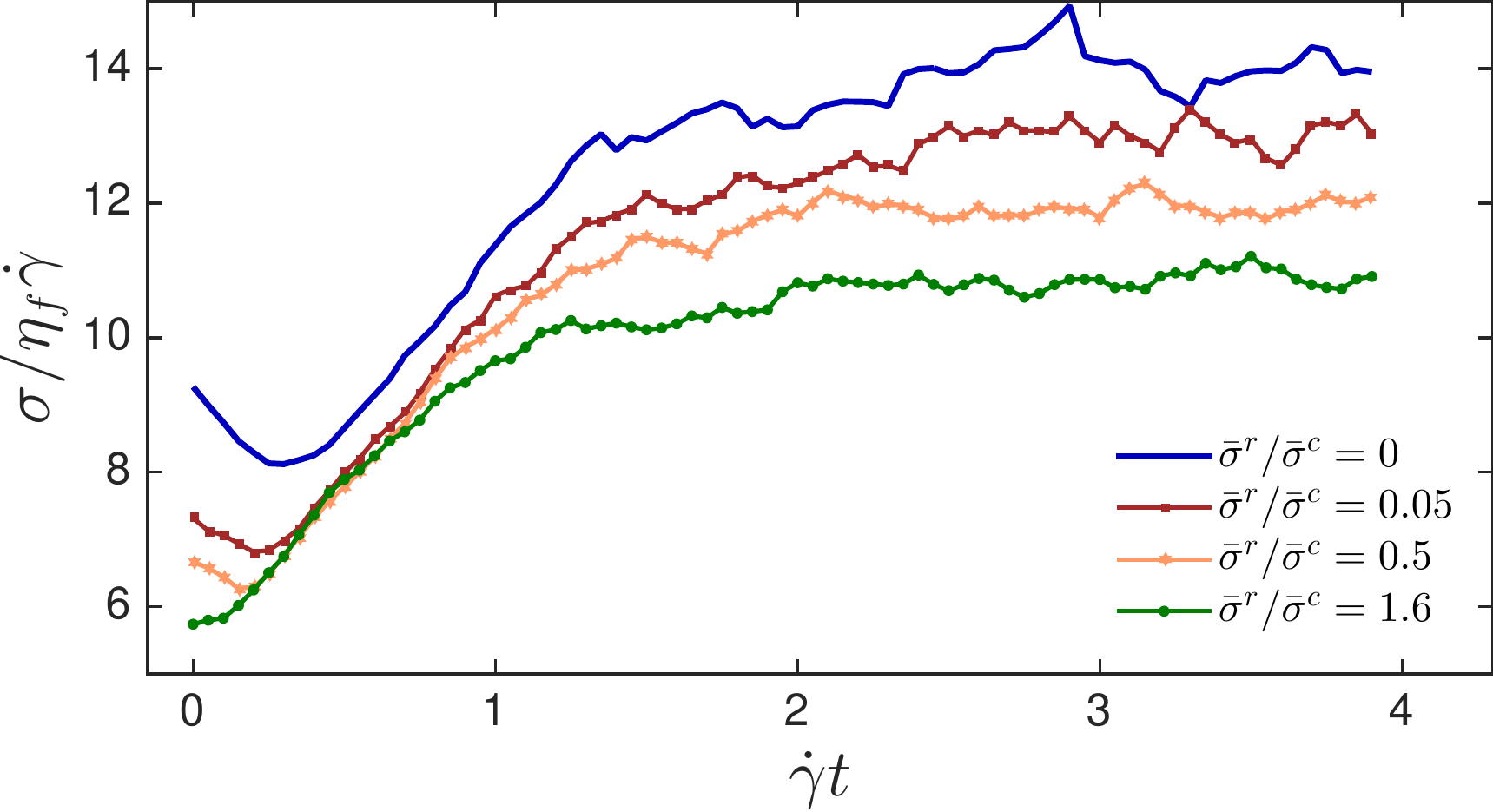}}

          \caption{
          Stress response to reversal as a function of
          (a) asperity length scale $h_\mathrm{min}$, with the numerical values of $h_\text{min}$ given in the legend in units of $d$;
          (b) friction coefficient $\mu_p$, with the numerical values of $\mu_p$ given;
          (c) repulsive force magnitude $|\mathbf{F}^r|$, quantified as the relative magnitude of the repulsive stress $\bar{\sigma}^r$ to the contact stress $\bar{\sigma}^c$.
          }
 \label{fig:roughness_and_hairs}
\end{figure}

In order to test the applicability of the above described mechanism to a wide range of particle systems, we address two factors pertaining to well studied suspensions, namely surface roughness and stabilising repulsion.
For suspensions of large particles (e.g., $d > 10 \ \mathrm{\mu m}$), such as the $40-50\mu m$ polystyrene spheres suspended in density matched silicon oils studied in~\cite{Acrivos1980}, surface roughness is perhaps the more relevant factor; for those of small particles (e.g., $d < 10 \ \mathrm{\mu m}$) steric or electrostatic repulsion may give well-defined repulsive forces.

\subsection{Surface roughness}

Surface roughness is represented numerically by an asperity length scale (by means of $h_\text{min}$) and the friction coefficient $\mu_p$. $h_\mathrm{min}$ contributes to the strain scale of the microfragile HS response and should also affect the HS magnitude. We explore such effects by varying $h_\mathrm{min}$ between $10^{-4}d$ and $10^{-2}d$, considering the physical size of surface asperities and bounded numerically by the singularity and the overdamping requirement at the lower and upper limits, respectively. The resulting $\sigma^l$ scaled by the steady state total stress $\bar{\sigma}$, following a reversal at $\dot{\gamma}t = 0$, is plotted against strain on a log-linear scale in Fig.~\ref{fig:roughness_and_hairs}(a). The strain scale of the microfragile peak decreases rather linearly with decreasing $h_\mathrm{min}$ in the $0.01d$ to $0.001d$ range, but saturates approaching $10^{-4}d$. We verified that the saturation is not due to inertial effects, but is perhaps due to the nonlinear coupling between particle configuration and dynamics. Decreasing $h_\mathrm{min}$ also significantly increases the peak magnitude, but only weakly affects the macrofragile HS response, the evolution of which interestingly collapses relative to $\bar{\sigma}$. The surface roughness effect of $h_\mathrm{min}$ thus controls the microfragile, but not the macrofragile HS response.

In a manner often employed in dry granular studies (see e.g.~\cite{Srebro2003}), we vary the friction coefficient $\mu_p$ incrementally between 0 and 1, exploring particle surfaces from ideally frictionless to very frictional. The total shear stress evolution is given on a linear $x$-scale in Fig.~\ref{fig:roughness_and_hairs}(b). On the contrary to the effects of varying $h_\mathrm{min}$, the microfragile response is largely insensitive to $\mu_p$, which is unsurprising given the small scale stress is dominated by attractive lubrication forces. The invariance of the hydrodynamic stress gives strong support to the central role of particle contacts in achieving the very different viscosities observed in such systems. The stresses differ hugely at larger strains, however, indicating that the $\mu_p$-dependent contact stress is important during $0.3<\dot{\gamma}t<2$, coinciding with recovery of $Z_c$ to its steady value. The increase of the contact stress with increasing friction can be understood from the increase of tangential contact forces and the decreased departure from the jamming volume fraction $\phi_c$ \cite{Liu1998}, which is known to decrease as friction increases~\cite{Chialvo2012,Sun2011}. The latter effect is also consistent with the experimental observation that the peak immediately after reversal becomes lower relative to the steady state stress when increasing volume fraction~\cite{Kolli2002a}. The interparticle friction thus mainly affects the large scale microstructure and contact stress and hence the macrofragile response. In reality, $\mu_p$ and $h_\mathrm{min}$ are probably simultaneously coupled to the surface roughness variation, though the combined effect may be deduced from the present separate analyses, exploiting the marked separation of scales associated with our two-scale description. 

\subsection{Surface stabilisation}

We next probe the effect of a generic stabilising repulsive force, extending the above analysis to consider particles in the size range $d<10\mu $m.
It is assumed, based on previous simulation results~\cite{Mari2014a,Seto2013}, that a static, short range, normal repulsive potential is sufficient to capture the essence of a stabilising mechanism such as electrostatic repulsion or a grafted polymer hair coating. Enhanced dissipation in the lubrication forces, a phenomena described by~\cite{Melrose2004}, is neglected for simplicity. 
A generic form of the repulsive force model derived by Fredrickson et al.~\cite{Fredrickson1991} is used,
$
\mathbf{F}^r = k \left(\frac{1}{h} \right)^{5/4} \mathbf{n}_{ij} \text{,}
$
where $k$ is some constant that encapsulates (among other things) the chemical properties of the hairs and their density on the surface, essentially quantifying the ``strength'' of the static repulsion.
We apply the same singularity regularisation as in the lubrication model, and the same values $h_\text{min}$, $h_{max}$. Coupling to the mechanical contact model is as before.
The total shear stress response to reversal is given in Fig.~\ref{fig:roughness_and_hairs}(c), for $k$ spanning 2 orders of magnitude (quantified by the relative magnitude of the steady-state repulsive $\bar{\sigma}^r$ and contact $\bar{\sigma}^c$ stresses). As expected for small $k$, the additional static repulsion is insufficient to separate particles, so the stress response closely resembles that for the base case in Fig.~\ref{fig:base_case}(a). As $k$ (or $\bar{\sigma}^r/\bar{\sigma}^c$) is increased, we note that while the large strain scale for the evolution appears to be unchanged, the steady value of $\sigma$ decreases. This is attributed to increasing inhibition to mechanical contacts (for which $h<0$) as the repulsion becomes stronger. We note that this trend is valid when $\bar{\sigma}^c$ is of comparable magnitude to $\bar{\sigma}^r$. For very large $\bar{\sigma}^r/\bar{\sigma}^c$, an opposite trend is observed~\cite{Mari2014a} due to a shear thinning mechanism---the polymer hair length can begin to contribute to an effectively larger total particle diameter, leading to a higher effective volume fraction and therefore a higher shear stress, as explained in detail by \cite{Mari2014a} and references therein. This then leads to shear thinning behaviour with \emph{reducing} $k$, rather than with \emph{increasing} $k$ as we observe here. For small strains after reversal, we observe a marked loss of the microfragile stress peak as $k$ is increased.

\begin{figure}
          \subfigure[]{
  \includegraphics[width=80mm]{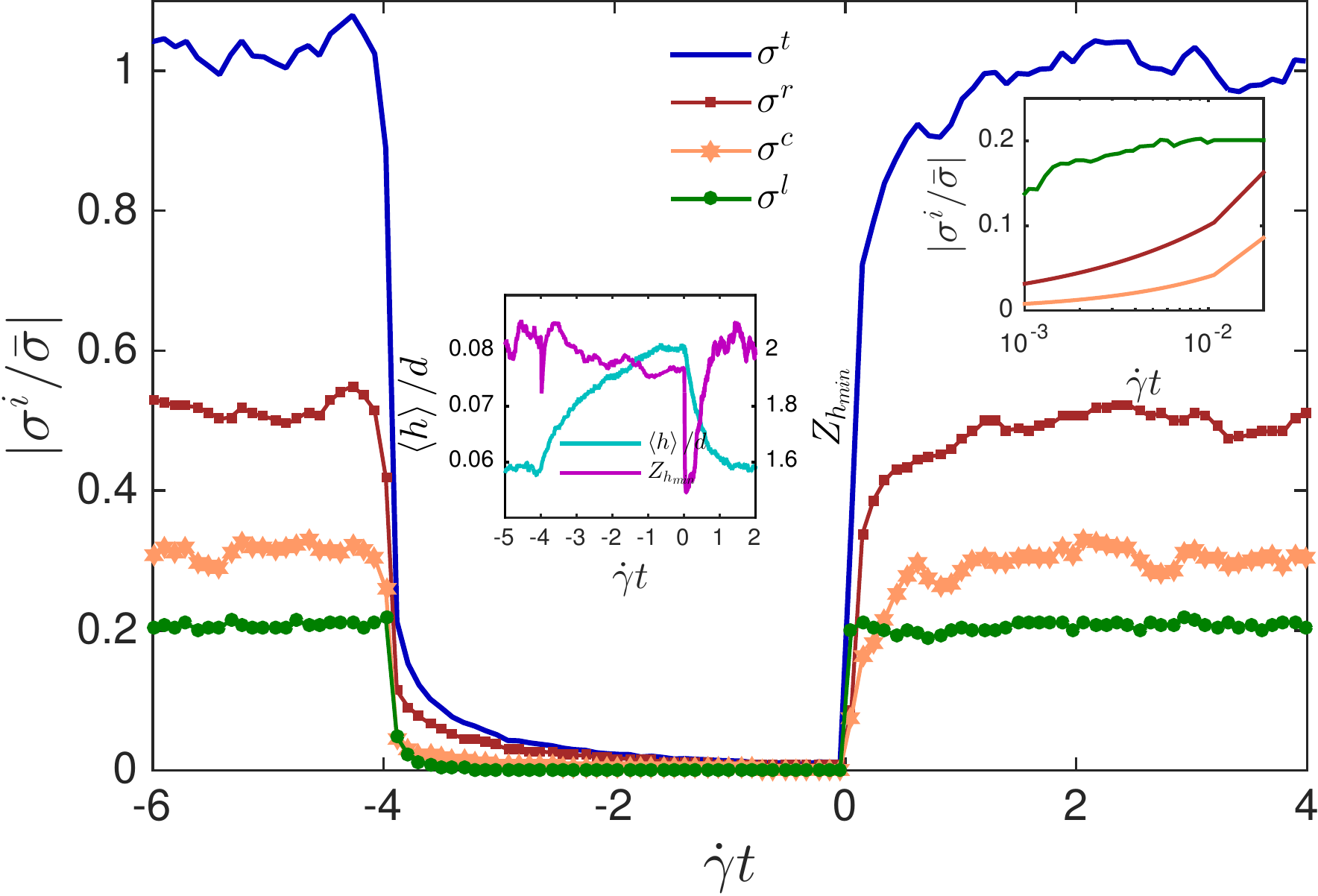}}
          \subfigure[]{
  \includegraphics[width=80mm]{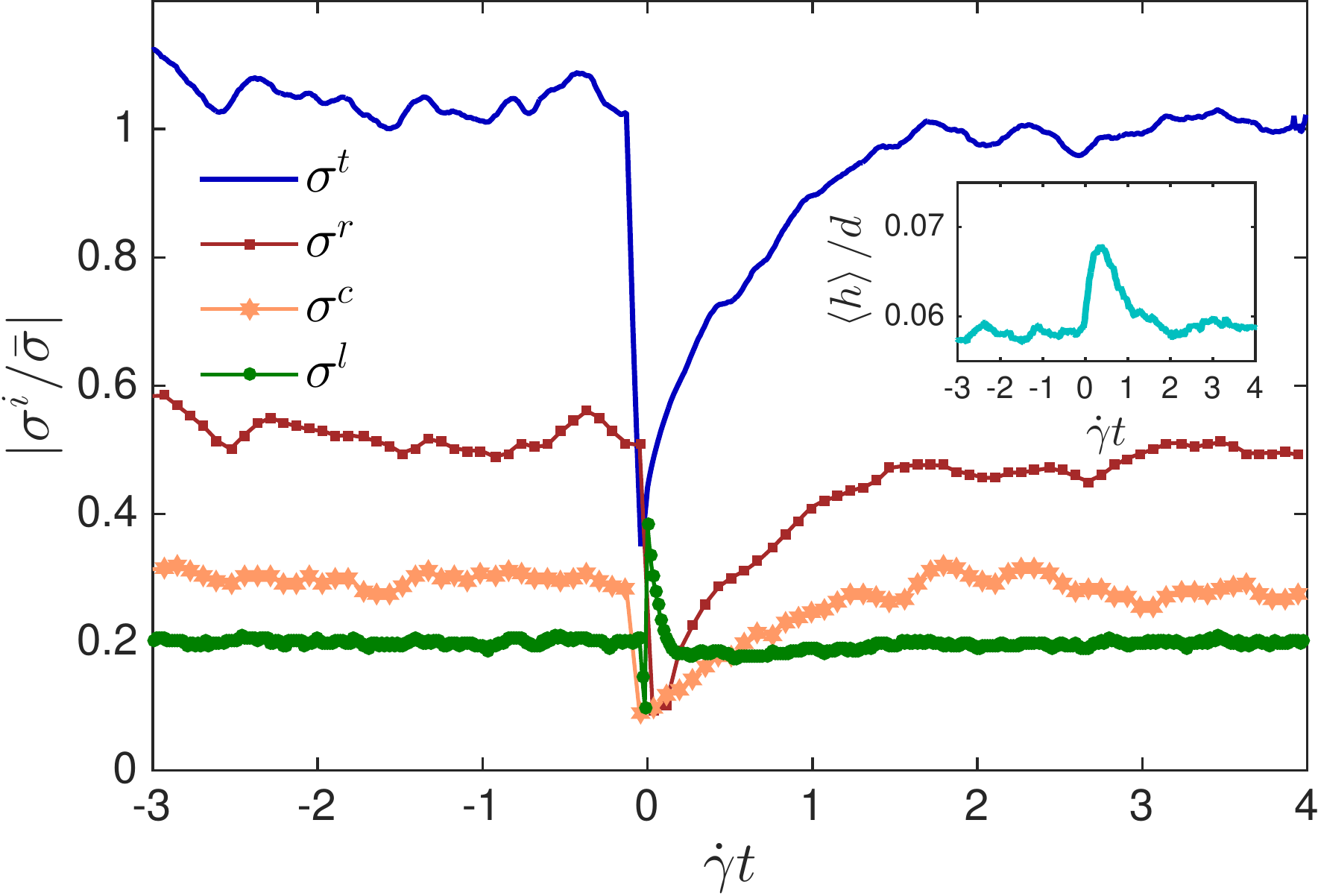}}
          \caption{
(a) Response to shear reversal, showing the total $\sigma$, contact $\sigma^c$, fluid $\sigma^l$ and repulsive stress $\sigma^r$ contributions, each scaled by the steady state total stress $\bar{\sigma}$;
      Right Inset: Same data with logarithmic $x$-axis.
      Middle Inset: Evolution of mean fluid film thickness $\left<h\right>$ and surface coordination $Z_\mathrm{h_{min}}$.
      (b) Analogous result for very short relaxation period.
      }
 \label{fig:base_case_rep}
\end{figure}

To further understand this loss, we present the full evolution of shear stress contributions for large $k$ and a flow cessation period sufficient to relax to steady state, Fig.~\ref{fig:base_case_rep}(a), with $\left<h\right>$ increasing from around $0.06d$ to $0.08d$ and $Z_{\mathrm{h_{min}}}$ decreasing modestly (middle Inset). To further characterise the relaxation period, we provide the associated $h$ distributions in Appendix~\ref{appendix:1}, under the action of the repulsive potential. Upon reversal, some remaining $h_\mathrm{min}$ contacts are opened, resulting in a $Z_{\mathrm{h_{min}}}$ decrease and a HS response over a $10^{-3}$ strain scale (middle and right Insets, respectively), consistent with the microfragile response in Fig.~\ref{fig:base_case}. In this sense, a microfragile HS response still occurs; though it starts from a ``loosened'' microstructure, producing a HS lower than its steady value, rather than the surge noted previously. Following this reasoning, a HS peak would be recovered if the relaxation period were shortened sufficiently to disallow any increase in $\left< h \right>$. We verify this in Fig~\ref{fig:base_case_rep}(b) using a very short relaxation period $\dot{\gamma}t = -0.01 \to 0$. A HS surge of about 100\% of its steady value is observed, although it does not result in an appreciable peak in the total stress since the HS contribution is small.  The repulsion and contact stresses similarly follow a macrofragile evolution. This again creates a strain window for the HS to be measured separately from other components. In short, the repulsive force magnitude together with the associated scales provides extra control over particle configurations and hence the stress response. The two-scale evolution concept is, however, still robustly helpful in understanding this more complicated behaviour.

\section{Concluding remarks}

We provide a robust characterisation of a two-scale response to shear reversal in dense suspensions, that is highly reminiscent of the micro- versus macro- fragility proposed by Cates~et~al.~\cite{Cates1998}.
Links are established between stress responses at small and large strains with \emph{microfragile} contact breakage and \emph{macrofragile} microstructural (re-)building respectively, resolving the hitherto unexplained nonmonotonic stress evolution following shear reversal. This substantiates the emerging understanding about the importance of particle contacts in suspension rheology -- they not only provide a significant contact stress at steady state, but also give rise to a pronounced small strain transient hydrodynamic response. 
This understanding provides a sound theoretical framework from which to formulate constitutive models with appropriate two-scale characteristics, and previous attempts at such models~\cite{Goddard2006} might be revised to correctly link the stress and microstructure at each scale. The evidence that different microstructural features control the contact and hydrodynamic stresses respectively and in an analogous way to that in dense granular flows~\cite{Sun2011}, supports further unification of dense suspension and granular rheology extending from steady~\cite{Boyer2011a} to unsteady state. The findings on surface features and interactions also open doors to either devising new experiments and protocols, e.g., varying relaxation time, to characterise particle surface properties and stress contributions; or designing new particles, e.g., with different grafted polymer hairs, to realise certain desired rheological properties. 


\section*{Acknowledgements}
This work is funded by the Engineering and Physical Sciences Research Council (EPSRC) UK and Johnson Matthey through a CASE studentship award. The authors would like to thank M. E. Cates and M. Hermes for their critical reading of this manuscript, and B. Guy, W. C. K. Poon, P. McGuire, M. Marigo, H. Stitt, H. Xu and J. Y. Ooi for helpful discussions.


    \appendix

\section{Further microstructural quantities for Figure 1}
\label{appendix:2}

\begin{figure}
\centering
      \subfigure[]{
  \includegraphics[height=22mm]{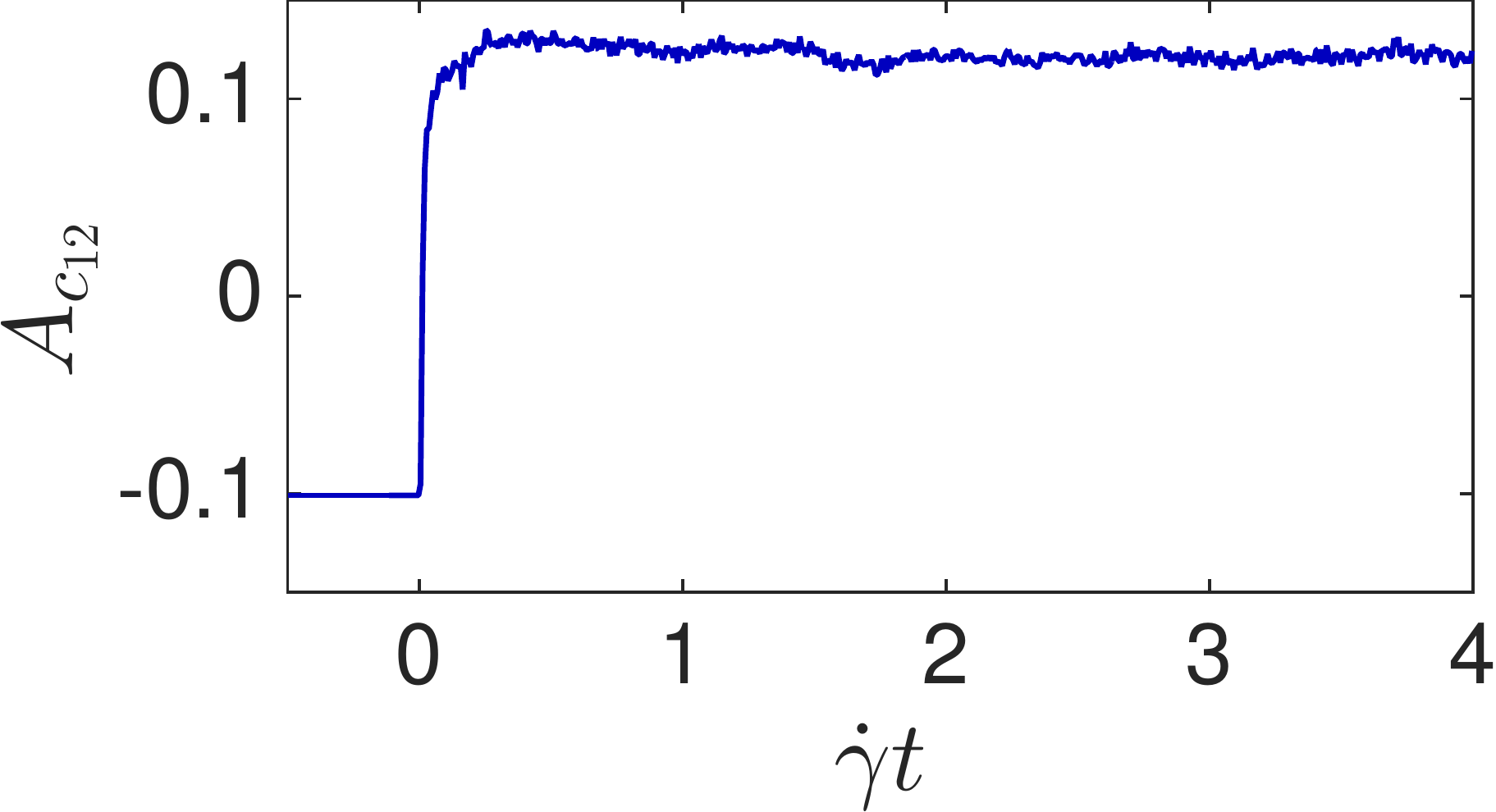}}
        \subfigure[]{
  \includegraphics[height=23mm]{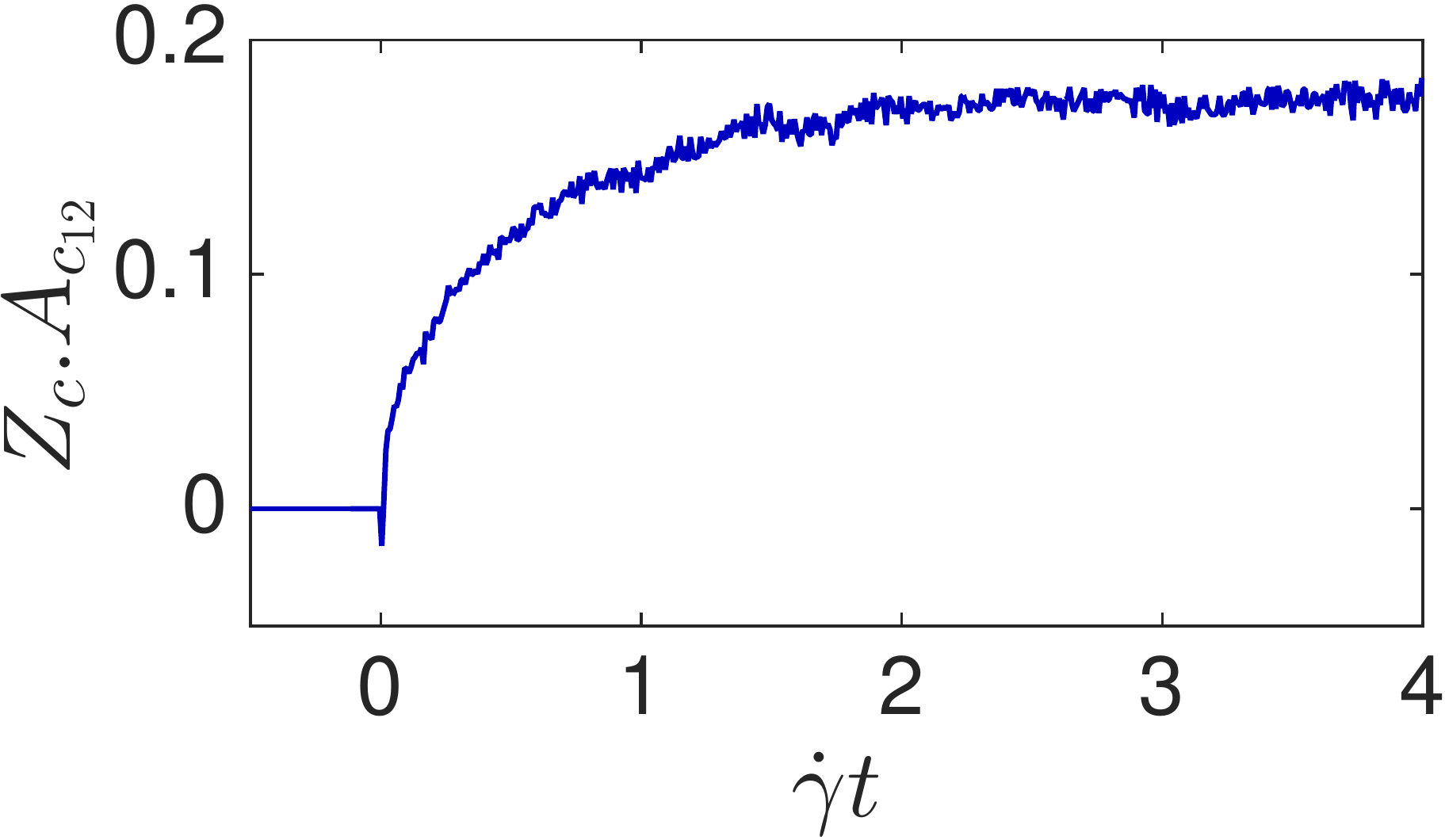}}
        \subfigure[]{
  \includegraphics[height=23mm]{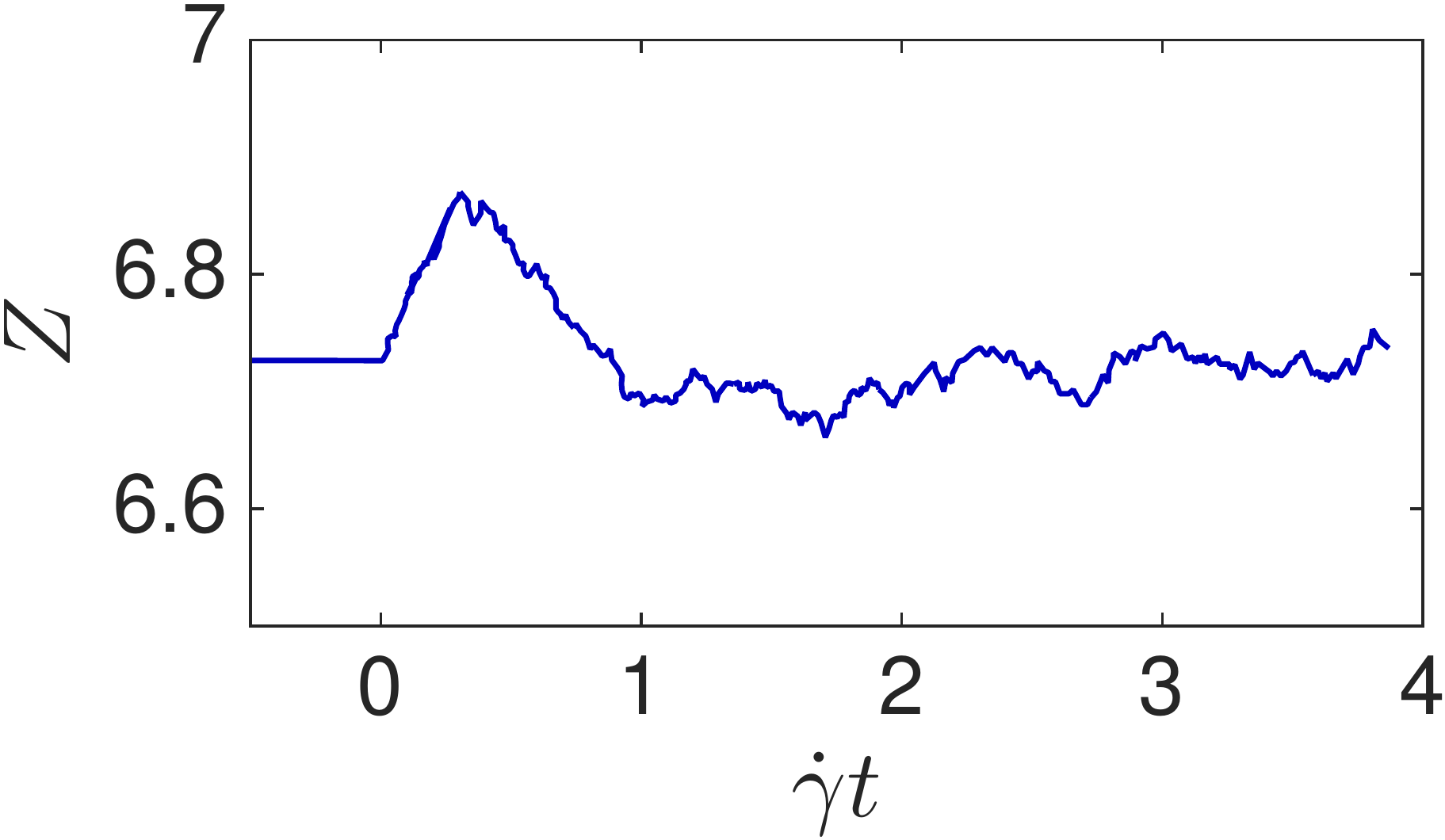}}
        \subfigure[]{
  \includegraphics[height=23mm]{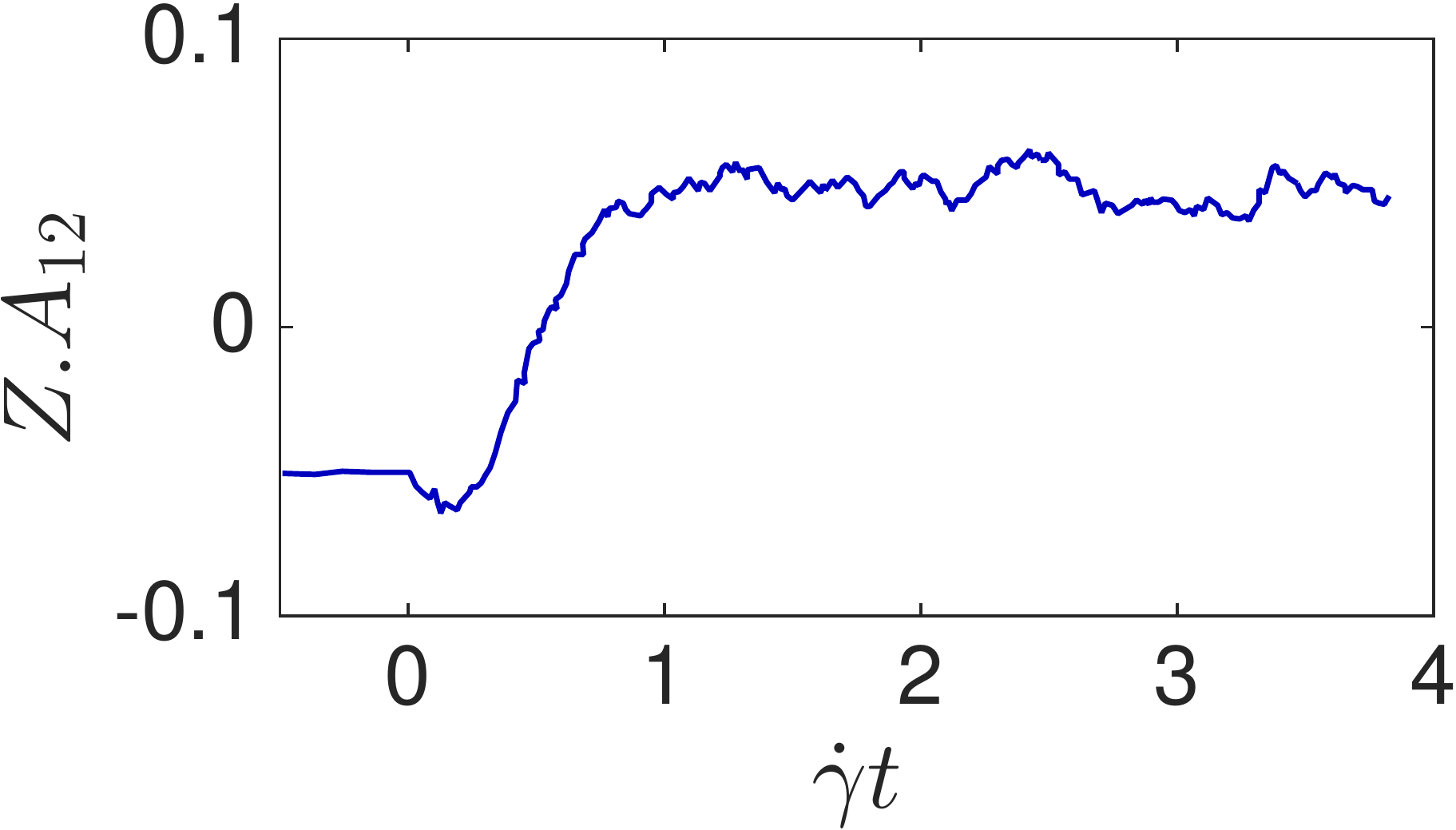}}
          \caption{
          Further microstructural quantities to support Figure 1(a). (a)-(b): Microstructural evolution evaluated at contact, to support $Z_c$; (c)-(d) Microstructural evolution evaluated at lubrication dutoff, to support $A_{12}$. Definitions of $Z$ and $A_{c_{12}}$ given in Appendix~\ref{appendix:2}, $Z_c$ and $A_{12}$ as defined in main article.
}
 \label{fig:base_case_addition}
\end{figure}

Additional microscopic quantities to back up the findings reported in Figure 1(a) of the main article are presented in Figure~\ref{fig:base_case_addition} of this document. $Z$ is a lubrication contact number, counting all pairs with $h<h_\text{max}$. All such pairs contribute to the fabric component $A_{12}$. $A_{c_{12}}$ is the shear component of the \emph{mechanical} fabric tensor, which we define as $\mathbf{A}_c=2/(Z_{c}N) \sum_{h<0} \mathbf{n}_{ij} \mathbf{n}_{ij}-\frac{1}{3}\mathbf{I}$, and omitting those pairs which support a contact stress less than $10^{-6}$ of the mean steady-state stress $\bar{P}$, consistent with the definition of $Z_c$ given in the main article.

\section{Distributions of $h$}
\label{appendix:1}
We plot the distribution of the particle-particle separation length $h$ for the simulations in Fig. 1(a) and Fig. 3 of the main article. It is noted that in the case with significant polymer hair repulsion, there remains a peak in $PDF(h)$ at very small $h$. This is consistent with the corresponding stress evolution, which demonstrates that there is still a non-negligible contribution from direct particle-particle contacts, $\sigma^c$.
\begin{figure}[H]
\centering
      \subfigure[]{
  \includegraphics[height=40mm]{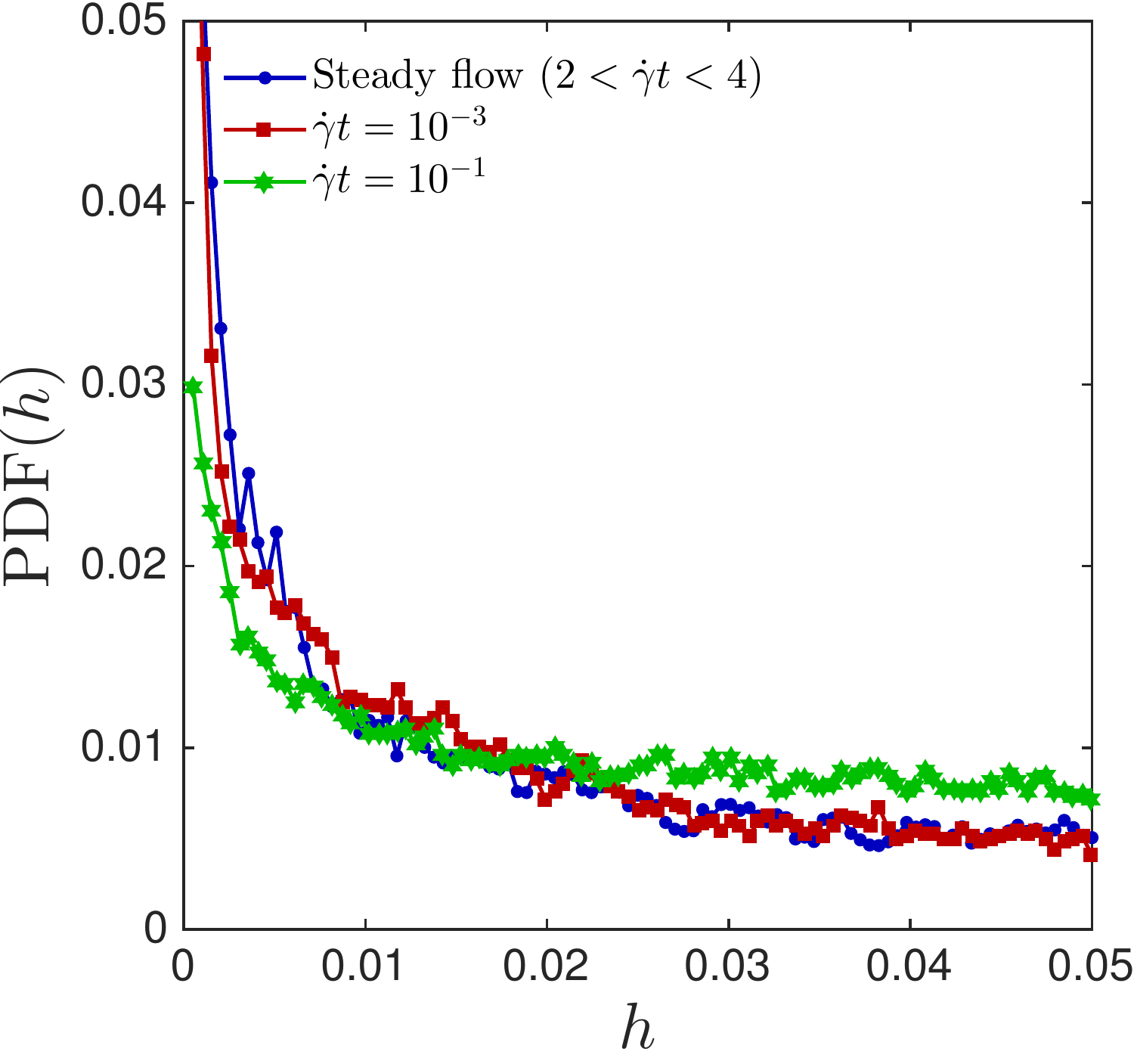}}
          \subfigure[]{
  \includegraphics[height=40mm]{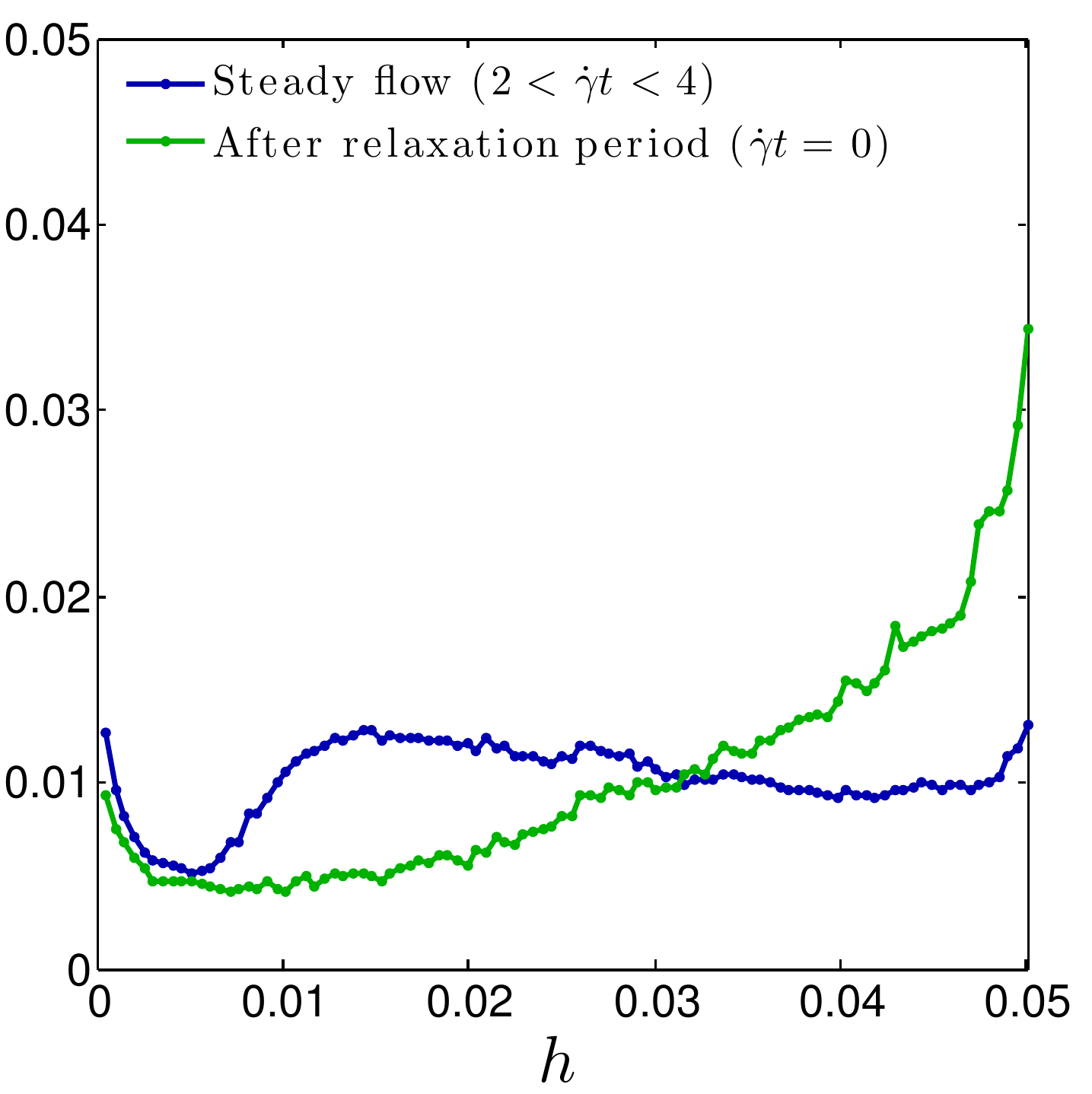}}
          \caption{
          Distribution of $h$ for the case in (a) Figure 1a and (b) Figure 3 of the article, focussing on the range for which lubrication forces are calculated.
}
 \label{fig:h_distribution}
\end{figure}

In addition, we find that the peak in $PDF(h)$ at small $h$ remains even after the relaxation period. We attribute this somewhat counter-intuitive finding to the repulsive force magnitude and cut-off scales and confinement effects. It is the subsequent opening of these remaining small $h$ particle pairs that is responsible for the very rapid evolution of $\sigma^l$ reported in Fig. 3 of the main article.

\bibliography{library.bib}
\bibliographystyle{ieeetr}

\end{document}